\input{aipcheck}

\documentclass[
    ,final            
  ]
  {aipproc}

\layoutstyle{6x9}

\begin{document}

\title{High Energy Observations of AGN Jets and Their Future Prospects}


\author{Jun KATAOKA}{
  address={Tokyo Institute of Technology, 2-12-1 Ohokayama, Meguro,
  Tokyo, 152-8551, JAPAN}
}

\begin{abstract}
In next five years, dramatic progress is anticipated for the AGN 
studies, as we have two important missions to observe celestial
 sources in the high energy regime: $GLAST$ and $Suzaku$.  
In this talk, I will summarize recent highlights in studies of AGN 
jets, focusing on the high-sensitivity X-ray observations 
that may shed new light on the forthcoming $GLAST$ era.  
I will especially present some examples from most recent $Suzaku$ 
observations of blazars, which provides important hints for the 
shock acceleration in sub-pc scale jets, as well as particle content 
in jets.  Then I will focus on the neutral iron-line feature observed 
in some broad line radio galaxies, as a probe of jet launching and/or 
the disk-jet connection. Finally, I will discuss new results of 
large scale (kpc to Mpc) jets recently resolved with $Chandra$ 
X-ray observatory. Simultaneous monitoring observations in 
various wavelengths will be particularly valuable for variable 
blazar sources,  allowing the cross correlations of time series 
as well as detailed modeling of the spectral evolution between the X-ray 
and gamma-ray energy bands. Possible impacts of these new observations 
across the electromagnetic spectrum on various spatial scales 
are discussed to challenge the long-standing mystery of AGN jet 
sources.
 
\end{abstract}

\maketitle


\section{1. Introduction}

Powerful, highly-collimated outflows called jets are commonly observed in a
wide variety of astronomical sources. It has been a long-standing 
mystery, however, where and how the relativistic jets are formed, 
and what is their composition.
From a theoretical standpoint, the relativistic AGN jets considered here
can be launched as outflows dominated by Poynting flux generated in the
force-free magnetospheres of black holes, or as hydromagnetic winds driven
centrifugally from accretion discs (see review by \cite{lov99}). 
In either case, strong magnetic
fields are involved in driving the outflows, although many (if not most)
observations indicate that eventually, {\sl particles}
carry the bulk of the jet's energy  (e.g., \cite{sik00}). 
This apparent discrepancy, however, can be resolved if the jets are 
indeed initially dominated by the Poynting flux but are 
efficiently converted into matter-dominated form at some later stage, 
most likely prior to the so-called ``blazar zone'' 
(see \cite{sik05} and references therein). Such a ``blazar
zone,'' the region where the bulk of the observed nonthermal radiation 
is produced, is most likely located at $r$ $\simeq$ 10$^3$$-$10$^4$ $r_g$,
where $r_g$ = $GM$/$c^2$ is the gravitational radius 
(\cite{spa00,kat01,tan03}).

This may indicate a scenario where a jet is launched near 
a rapidly rotating black hole, presumably at the innermost 
portions of the accretion disk (see, e.g., \cite{koi99}). 
Such a jet, initially consisting of protons and electrons, 
is accelerated by large scale magnetic field stresses and 
within 100 $r_g$ can be loaded  by electron/positron 
($e^{-}e^{+}$) pairs via interactions with the coronal 
soft gamma-ray photons (note that such photons are directly 
seen in the spectra of Seyfert galaxies; see, 
e.g., \cite{zdz00}). Hence, it is possible 
that relativistic jets in quasars (beyond the jet formation zone) 
may well contain more electron/positron pairs than protons, 
but are still dynamically dominated by cold  protons. 
Apparently, observational
evidence is strongly awaited to test this hypothesis, which provides a 
direct hint for the connection between the jet and accretion
disk at the very inner-site ($\le$ 1 mpc scale) of the jet.

Meanwhile, extragalactic jets constitute the longest collimated 
structures in the Universe. They transport huge amounts of energy 
from the nuclei of active galaxies out to kpc or Mpc distances, 
significantly affecting the properties of the surrounding 
intracluster/intergalactic medium. These large scale 
jets have been extensively 
studied in the radio domain on different scales since the very
beginning of the development of modern radio interferometers 
(e.g., \cite{beg87}). More recently, the excellent spatial
resolution of the {\it Chandra} X-ray Observatory (and, to a lesser extent,
of other X-ray satellites like {\it XMM-Newton}) has allowed us to
image large-scale structures in powerful extragalactic radio sources
at X-ray frequencies as well, and thus has opened a new era in studying the
high energy emission of these objects. More than 100 radio-loud AGNs are
now known to possess X-ray counterparts to their radio jets, hotspots or
lobes on kpc-to-Mpc scales (e.g., \cite{har02,har06,sam04,kat05} 
and references therein). 
Here we briefly overview recent highlights from the X-ray/gamma-ray 
observations of AGN jets on various scales from 1 mpc to 1Mpc, as a 
new challenge to the jet physics.

\section{2. sub-pc jets: Blazars}

Observations with the EGRET instrument on board the {\it Compton 
Gamma-Ray Observatory} in the gamma-ray band have opened a new window
for studying AGN jets, and revealed that many radio-bright and variable
AGN are also the brightest extragalactic MeV$-$GeV gamma-ray emitters
(see, e.g., \cite{har99}).  The properties of the gamma-ray
emission in those objects --- often termed ``blazars'' --- supported earlier
inferences based on radio and optical data, and independently indicated
significant Doppler boosting, implying the origin of broad-band emission
in a compact, relativistic jet pointing close to our line of sight.
Generally, the overall spectra of blazar sources (plotted in
the log$(\nu)$-log$(\nu F_{\nu})$ plane, where $F_{\nu}$ is
the observed spectral flux energy density) have two pronounced continuum
components: one peaking between IR and X-rays and
the other in the gamma-ray regime (see, e.g., \cite{kub98,ghi98}). 
The lower energy component is believed to be produced
by the synchrotron radiation of relativistic electrons accelerated
within the outflow, while inverse Compton (IC) emission by the same
electrons is most likely responsible for the formation of the high energy
gamma-ray component.  

It is widely believed, in addition, that the IC emission from quasar 
hosted blazars (QHBs) is dominated by the scattering of 
soft photons external to the jet (external Compton process, ERC), 
which are produced by the accretion disk, either directly or 
indirectly via scattering/reprocessing in the broad line region 
(BLR) or dusty torus (see, e.g., \cite{sik94}). 
Other sources of seed photons can also contribute to the 
observed IC component, in particular the synchrotron photons 
themselves via the synchrotron self-Compton process (SSC) which 
is often the case for BL Lacertae objects (e.g., \cite{kub98}).
In some cases, gamma-ray emissions seen to extend to the 
TeV range; the X-ray and TeV gamma-ray bands corresponds to the 
highest energy end of the synchrotron/IC emission. As we see below, 
detailed modeling of  broad-band blazar emission as well as 
temporal variability can provide information about the location of
 the dissipative regions in blazars, the energy distribution 
of relativistic electrons/positrons, the magnetic field intensity, 
and the jet power.

\subsection{2.1 1ES 1218+304 (HBL)}

1ES 1218+304 is categorized as a high-frequency BL Lac object (HBL), 
as a redshift $z$ = 0.182. It was discovered as a TeV emitter by MAGIC 
at energies $\ge$ 100 GeV (\cite{alb07}) and subsequently
confirmed by VERITAS (\cite{for07}). The source was observed with 
$Suzaku$ during 2006 May 20-21 UT, yielding a net exposure time of 79.9 
ks (\cite{sat08}). Figure 1 shows the averaged light curves of the four XISs in
the lowest (0.3$-$1.0 keV) and the highest (5$-$10 keV) 
X-ray energy bands. Interestingly, the observed flare shows the 
following characteristics: (1) The flare shape is asymmetric in time 
($t_{\rm r}$/$t_{\rm d}$ $\le$ 1) especially in the lower energy band 
(but note  $t_{\rm r}$/$t_{\rm d}$ $\simeq$ 1  for 5$-$10 keV light
curve). (2) The flare amplitude becomes larger as the photon energy 
increases. (3) The risetime of the flare is almost constant below 2 keV,
while it becomes gradually longer at higher energy bands.
In this context, we try to evaluate lags of temporal variations in various 
energy bands. We found that the hard X-ray
(5$-$10 keV) peak lagged behind that in the soft X-ray 
(0.3$-$1 keV) by (2.3$\pm$0.7)$\times$10$^4$ sec. 

\begin{figure}
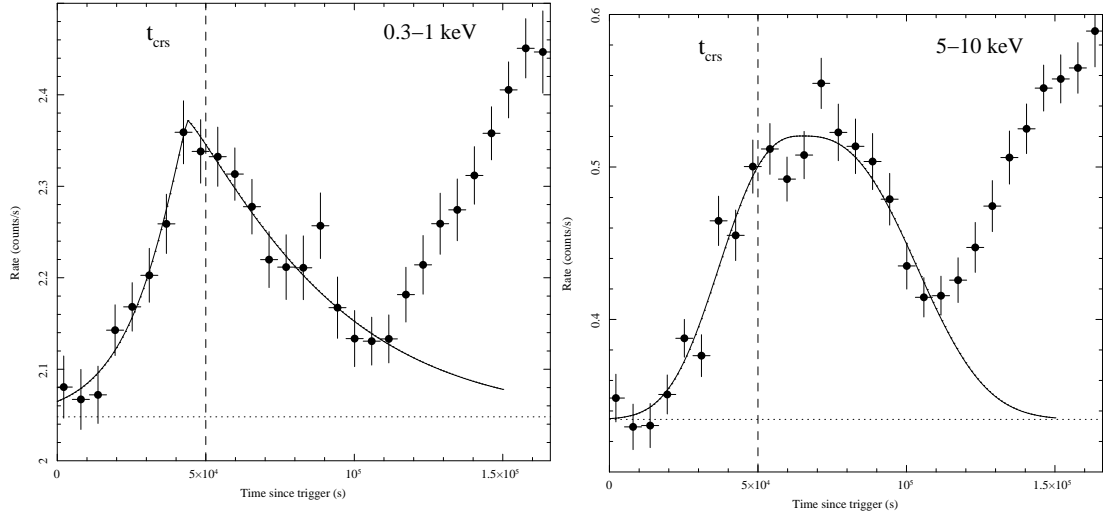

  \includegraphics[height=.33\textheight,angle=270]{0.3-1_model_2880s.ps}
  \includegraphics[height=.33\textheight,angle=270]{5-10_model_2880s.ps}
  \caption{$Suzaku$ XIS Light curves in the two energy bands:
0.3$-$1 keV($left$) and 5$-$10 keV ($right$). The dotted line shows 
the best-fit model presented in \cite{sat08}.}
\end{figure}

This is completely opposite to a well-known ``soft-lag'', as has 
been obtained from the past observations (e.g., \cite{tak96,kat00}). 
In the theoretical context, however, hard-lag 
is actually expected especially in the X-ray variability of TeV blazars, 
but has never been observed so clearly before. It has been suggested
that a hard-lag is observable only at energies closer
to the maximum electron energy (\cite{kir98}), where the acceleration time is almost comparable
to the cooling time scale of radiating electrons: $t_{\rm acc}(E_{\rm max})$
$\simeq$ $t_{\rm cool}(E_{\rm max})$. Noting that the typical synchrotron 
emission frequency, averaged over pitch angles, of an electron with
energy $\gamma$$m_e c^2$ is given by  $\nu$ $\sim$
3.7$\times$10$^6$$B$$\gamma^2$ Hz, we obtain; 
\begin{eqnarray*}
t_{\rm acc} (E)  &=& 9.65\times10^{-2} (1+z)^{3/2} \xi B^{-3/2} \delta^{-3/2} E^{1/2} {\rm s},\\
t_{\rm cool} (E) &=& 3.04\times10^{+3} (1+z)^{1/2} B^{-3/2} \delta^{-1/2} E^{-1/2} {\rm s},
\end{eqnarray*}
where $E$ is the observed photon energy in unit of keV, $z$ is the redshift, 
$B$ is the magnetic field strength, and $\delta$ is the beaming factor. 
Note, for lower energy photons ($E$ $\ll$ 
$E_{\rm max}$), $t_{\rm acc}(E)$ is always shorter 
than $t_{\rm cool}(E)$ because  higher energy electrons need longer time to 
be accelerated 
($t_{\rm acc}(\gamma)$ $\propto$ $\gamma$) but cools rapidly 
($t_{\rm cool}(\gamma)$ $\propto$ $\gamma^{-1}$).  This energy
dependence of acceleration/cooling time-scales may qualitatively 
explain observed characteristics of X-ray light curves of 1ES~1218+304.

\begin{figure}
  \includegraphics[height=.35\textheight]{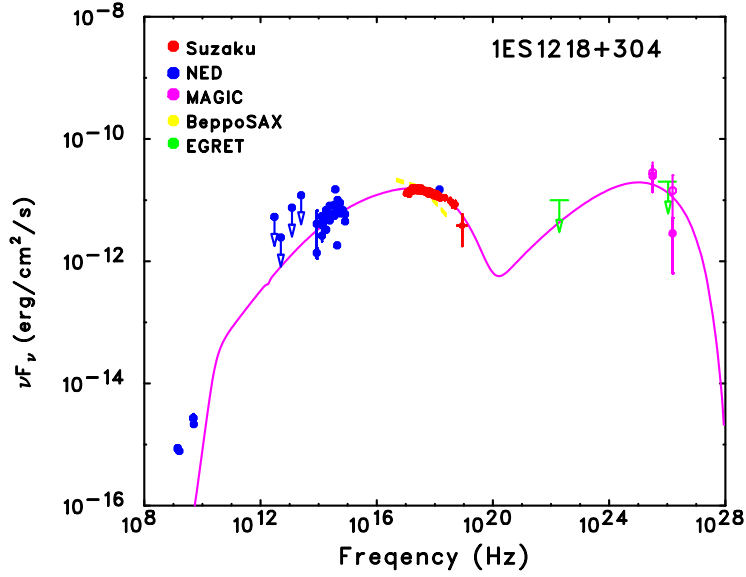}
  \caption{SED of 1ES 1218+304. The solid line shows a one-zone SSC
 model assuming the parameters: $B$ = 0.047 G, $\delta$ = 20, 
$s$ = 1.7, $\gamma_{\rm  min}$ = 1, $\gamma_{\rm brk}$ = 
8$\times$10$^3$ and $\gamma_{\rm max}$
 = 8$\times$10$^5$, where $s$ is the electron spectral index. See
 \cite{sat08} for more detail.}
\end{figure}

It is thus interesting to consider a simple toy model 
in which the rise time of the flare is primarily
controlled by the acceleration time of the electrons corresponding
to observed photon energies, while the fall time of
the flare is due to the synchrotron cooling time scale. In this
model, the amount of hard-lag is simply due to the difference
of $t_{\rm acc}$, and independent of the energy dependence of
$t_{\rm cool}$:
\begin{equation}
\tau_{\rm lag} = E_{\rm acc} (E_{\rm hi}) - E_{\rm acc} (E_{\rm low}) 
\sim 9.65\times10^{-2} (1+z)^{3/2} \xi B^{-3/2} \delta^{-3/2} (E_{\rm hi}^{1/2} 
- E_{\rm low}^{1/2}) \hspace{3mm}{\rm s},\\
\end{equation}
where $E_{\rm low}$ and $E_{\rm hi}$ are the lower and higher 
X-ray photon energies to which the time-lag is observed. 
Assuming $\delta$ = 20 from multiband spectral fitting (see Figure 2), 
the best fit 
parameter of the magnetic field $B$ can be written as $\simeq$ 0.05
$\xi_5$ G, where $\xi_5$ is the gyro-factor in units of
10$^5$. As discussed in detail in \cite{sat08}, the 
above toy model qualitatively well represents  
the observed spectral/temporal features of 1ES 1218+304, in particular: 
(1) the synchrotron component peaks around the $Suzaku$ XIS 
energy band in the multiband spectrum and (2) the observed 
light curve is symmetric in shape when measured at the high energy band,
while being asymmetric at the lower energy band.

\subsection{2.2 PKS 1510-089 (QHB)}

In contrast to the HBLs as presented above, 
the X-ray band corresponds to the low-energy end of the inverse 
Compton emission for most QHB-type blazars. For these sources, 
a probe of the low energy electron/positron content in blazars 
was proposed by \cite{begs87}, 
and extensively studied in the literature (\cite{sik00,mod04,cel07}). 
The gamma-ray emission 
is produced by electrons/positrons 
accelerated {\it in situ}, and thus
before reaching the blazar dissipative site the
electrons/positrons are expected to be cold. If they are transported
by a jet with a bulk Lorentz factor
$\Gamma_{\rm jet} \ge 10$, they upscatter external UV 
photons up to X-ray energies
and produce a relatively narrow feature expected to be located
in the soft/mid X-ray band, with
the flux level reflecting the amount of cold electrons and
the jet velocity.  Unfortunately, such an additional
bulk-Compton (BC) spectral component is difficult to observe because
of the presence of strong non-thermal blazar emission, which dilutes
any other radiative signatures of the active nucleus. In this context,
QHBs may constitute a possible exception, since their
non-thermal X-ray emission is relatively weak when compared to other
types of blazar sources.

PKS~1510$-$089 is a nearby ($z$ = 0.361) QHB 
detected in the MeV$-$GeV band by EGRET. It is a highly
superluminal jet source, with apparent velocities of $v_{\rm app}$
$\ge$ 10 $c$ observed in multi-epoch VLBA observations
(e.g., \cite{hom01}).
Recent observations by {\it BeppoSAX} (\cite{tav00}) 
confirmed the presence of a soft X-ray excess below 1\,keV, that 
may be among the best candidates for detecting the BC bump.
The observational campaign of PKS 1510-089 in 2006 August 
commenced with a deep $Suzaku$ observation lasting three days for a total exposure 
time of 120\,ks, and continued with $Swift$ monitoring over 18 days 
(\cite{kat08}). Besides $Swift$ observations, which sampled 
the optical/UV flux in all 6 UVOT filters as well as the X-ray spectrum 
in the 0.3$-$10\,keV energy range, the campaign included ground-based optical
and radio data, and yielded a quasi-simultaneous broad-band spectral energy
distribution from 10$^9$\,Hz to 10$^{19}$\,Hz.

Figure 3 ($left$) shows an overall SED of PKS 1510-089, whereas 
Figure 3 ($right$) shows  in detail the optical--to--X-ray region of
the SED. In the $right$ panel, 
the hump on the left mimic an excess emission from the dusty
torus as suggested by {\it IRAS} with a dust
temperature of $kT$ $\simeq$ 0.2\,eV and $L_{\rm dust}$ $\simeq$
3.7$\times$10$^{45}$\,erg\,s$^{-1}$.
The hump on the middle is our attempt to account for the blue bump
assuming an inner-disk temperature of $kT$ $\simeq$ 13\,eV and
$L_{\rm disk}$ $\simeq$ 4$\times$10$^{45}$\,erg\,s$^{-1}$.
From the spectral fitting of the $Suzaku$ data, we found that
the 0.3$-$50 keV spectrum is well represented by an extremely 
hard power-law with photon index $\Gamma$ = 1.2, augmented by 
a black-body--type emission of $kT$ $\simeq$ 0.2\,keV (\cite{kat08}).  
Figure 4  shows count rate variations during the
$Suzaku$ observation. The time variation of underlying power-law
component and the soft X-ray excess are separately shown in this figure.
This clearly indicates different variability properties: count rates
only slightly decreased for PL  component, while it reached
a delayed maximum $\sim$1.5 day from the start of the
$Suzaku$ observation for the soft X-ray hump. 

\begin{figure}
  \includegraphics[height=.32\textheight]{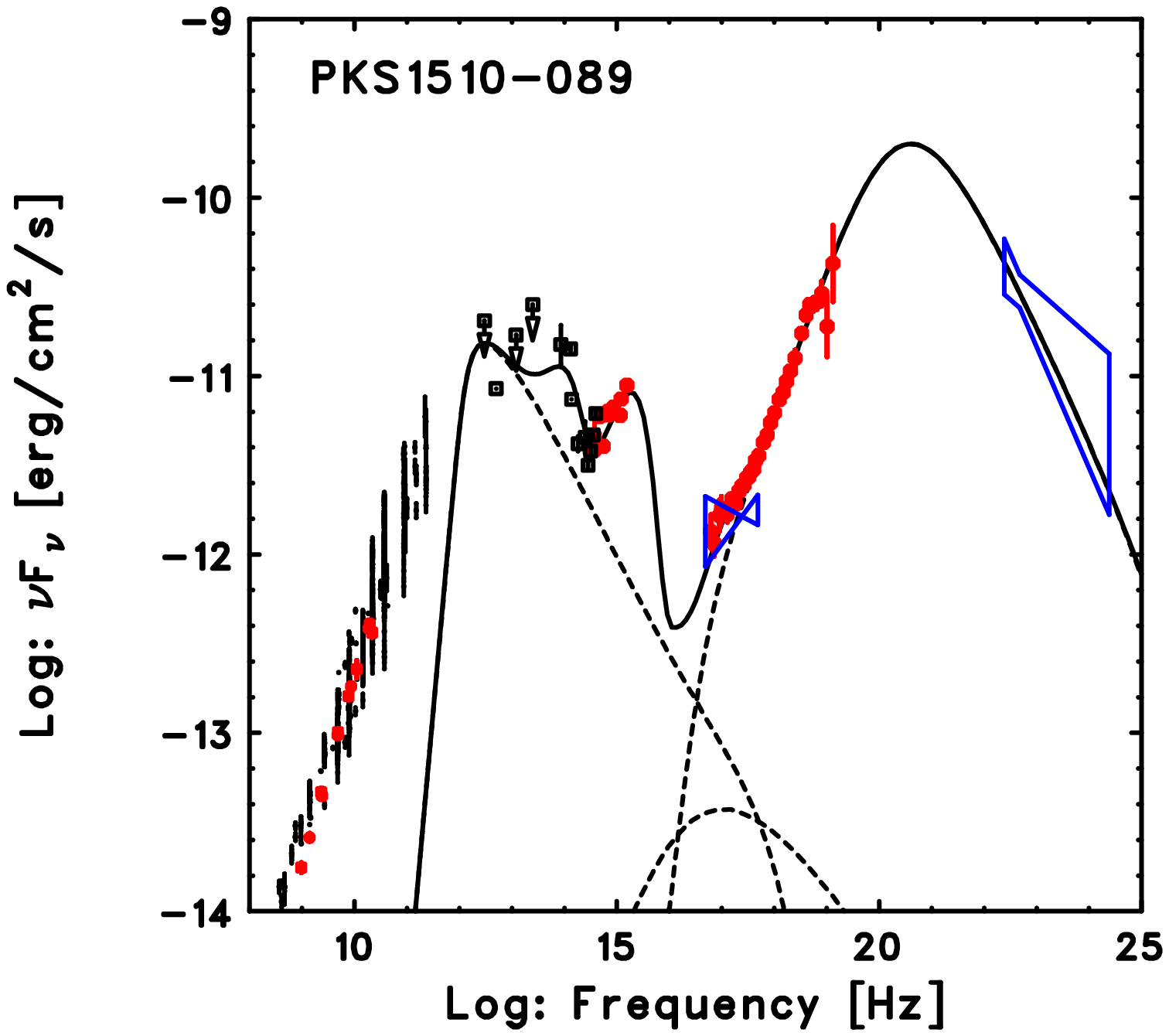}
  \includegraphics[height=.32\textheight]{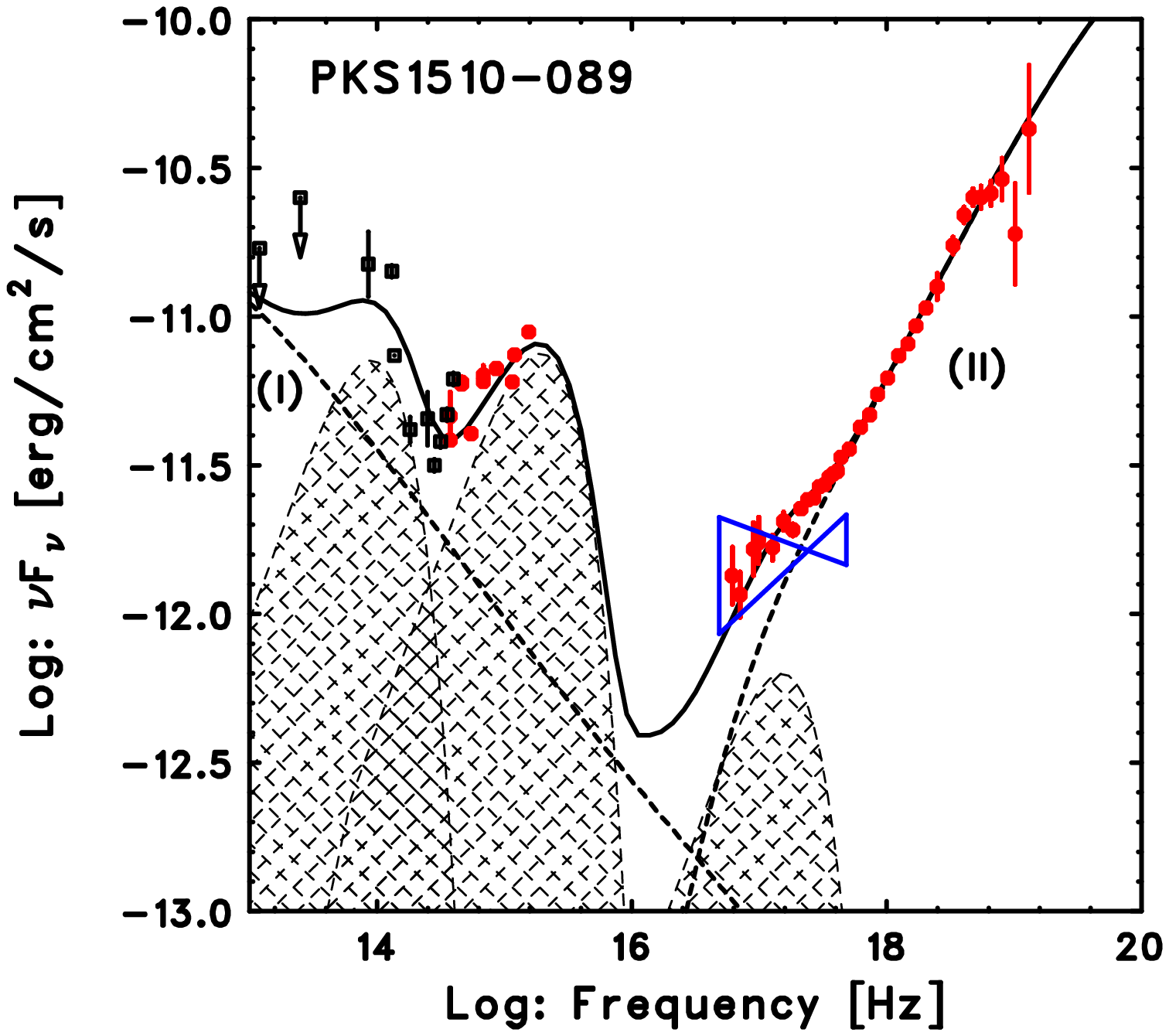}
  \caption{$left$: Overall SED of PKS 1510-089 constructed with
 multiband data obtained during 2006 campaign. $right$: A close-up of
 the SED between the optical and X-ray bands. See \cite{kat08}  
for more details.}
\end{figure}

We investigate below whether such excess can be produced by a bulk 
Comptonization of external diffuse radiation by
cold inhomogeneities (with bulk Lorentz factors $\Gamma_1$ and
$\Gamma_2$) and/or density enhancements prior to their collisions.
At $r > r_{\rm BLR}$, where $r_{\rm BLR}$ is the distance of the 
broad line region from the nucleus, density of the diffuse external 
UV radiation is very small, while bulk-Compton features from upscatterings of
dust infrared radiation falls into the invisible extreme-UV band.
However, if acceleration of a jet has already occurred at 
$r \le r_{\rm BLR}$, upscattering of photons from broad-emission 
line region should lead to
formation of bulk Compton features, with peaks located around
$\nu_{\rm BC,i} \sim {\cal D}_{\rm i} \Gamma_{\rm i} \nu_{\rm UV}/(1+z)$ and
luminosities
\begin{equation} L_{\rm BC,i} = {4 \over 3} c \sigma_{\rm T} u_{\rm BLR} \Gamma_{\rm i}^2
{\cal D}_{\rm i}^4 N_{e,{\rm obs, i}} \, ,
\end{equation}
where $i=1,2$, $u_{\rm BLR}$ is the energy density of the broad emission lines,
${\cal D}_{\rm i}$ is the Doppler factor, and  $N_{e,{\rm obs, i}}$ 
is the number
of electrons and positrons contributing
to the bulk-Compton radiation at a given instant (see \cite{mod04}).
For the conical jets the Doppler factor should  be replaced by the 'effective'
Doppler factor which for $\theta_{\rm obs} \le \theta_{\rm jet}$ is
${\cal D}_{\rm i} = \kappa \Gamma_{\rm i}$, where $1< \kappa <2$.
For our model parameters
\begin{equation}
N_{e,{\rm obs,1}} \simeq {N_{\rm inj} \over 2}  \,
{r_{\rm BLR} \over
\lambda_0 {\cal D}_{1}} \, , 
\end{equation}
and
\begin{equation}
N_{e,{\rm obs,2}} \simeq {N_{\rm inj} \over 2}  \,
{r_{\rm BLR} \over
\lambda_0 {\cal D}_{2}} \,
{\Gamma_{\rm sh}^2 \over 2\Gamma_2^2}
\, ,
\end{equation}
where $\lambda_0$ is the proper width (longitudinal size) of the cold
inhomogeneities (see Appendix A3 in \cite{mod04}). 
With the above  approximations and $\kappa = 1.5$ our model predicts
location of the bulk-Compton features at $\sim 1$ keV and $\sim 18$ keV,
and luminosities of $\sim 2 \times 10^{44}\, {\rm erg \, s}^{-1}$
and $2 \times 10^{46}\,{\rm erg \, s}^{-1}$, respectively.
Thus it seems that within the uncertainties regarding the details
of the jet geometry and model parameters,
bulk-Compton radiation produced by slower
inhomogeneities is sufficiently luminous to be responsible
for the soft X-ray excess observed by $Suzaku$, while the faster one
can be tentatively identified with a small excess at $ \sim 18$ keV
seen in Figure 3 ($right$).

\begin{figure}
  \includegraphics[height=.35\textheight]{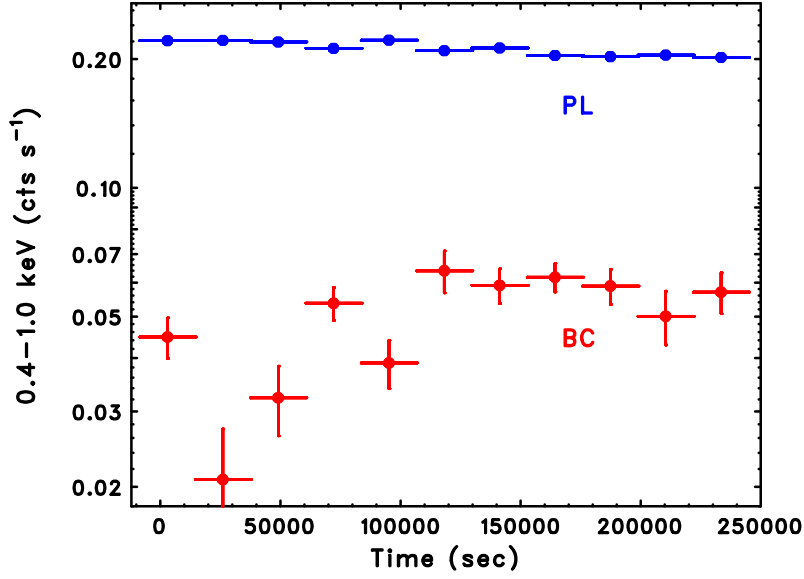}
  \caption{The variability of PKS 1510-089 observed with $Suzaku$. While
 the underlying power-law component is stable, soft X-ray excess is
 highly variable with a delayed maximum.}
\end{figure}

Finally, we shortly comment on the pair content of a jet.
In our SED modeling (Figure 3 $left$). the amount of electrons/positrons 
injected into shell by the end of the shock operation is
\begin{equation} 
N_{e,{\rm inj}} = t_{\rm sh}' \int_{\gamma_{\rm min}} Q_{\gamma} d\gamma \simeq
{\Delta r \over c \Gamma_{\rm sh}} {K_e \over (p-1) \gamma_{\rm min}^{p-1}} \simeq
2.9 \times 10^{53} \, ,
\end{equation}
where $t_{\rm sh}' = \Delta r/(c\Gamma_{\rm sh})$ is the lifetime of the shock
as measured in the shock (discontinuity surface) rest frame.  
The electrons/positrons are accelerated/injected resulting in an 
average energy $\bar \gamma_{\rm inj} = \int Q_{\gamma} \gamma d\gamma / \int Q_{\gamma} d\gamma
\simeq 22$. Assuming that this energy is taken from protons, we have
the electron$+$positron to proton ratio 
\begin{equation}
{N_e \over N_p} = \eta_e {m_p (\bar\gamma_p -1) \over m_e
\bar\gamma_{\rm inj}} \, ,
\end{equation}
where $\eta_e$ is the fraction of the proton thermal energy 
tapped by electrons and positrons.
The value of $\bar\gamma_p-1$, which actually represents efficiency of the 
energy dissipation, depends on properties and speeds of colliding 
inhomogeneities, and is largest if they have   same rest densities and  masses.
In this case, assuming $\Gamma_2 > \Gamma_1 \gg 1$,  
\begin{equation} 
\bar\gamma_p-1 =
{(\sqrt{\Gamma_2/\Gamma_1}-1)^2 \over 2 \sqrt{\Gamma_2/\Gamma_1}} \, ,
\end{equation}
and $\Gamma_{\rm sh}=\sqrt{\Gamma_1 \Gamma_2}$ (\cite{mod04}).
For reasonable choice of $\Gamma_1 = 10$ and $\Gamma_2 = 40$, 
we obtain $N_e/N_p \sim 20 \eta_e$. The 
result supports our original hypothesis that the power of the jet is 
dominated by protons but wit a number of electrons/positrons 
exceeding a number of protons by a factor $\sim$ 10 (\cite{kat08}).

\section{3. Disk-Jet Connection: 3C~120}

All AGNs are thought to be powered by accretion of matter onto a 
supermassive black hole, presumably via an equatorial 
accretion disk. Recent VLBI observations of a nearby 
active galaxy M87 confirmed that the jet is $already$ launched 
within $\sim$60 $r_g$ (where $r_g$ = $GM$/c$^2$ is the gravitational radius), 
with a strong collimation occurring within $\sim$200 $r_g$ 
of the central black hole (\cite{jun99}). 
These results are consistent with the hypothesis that jets are formed 
by an accretion disk, which is threaded by a magnetic field. 
Therefore the observational properties of the accretion disk 
and corona are essential ingredients to jet formation 
(e.g., \cite{liv99} and references therein).

In this meaning, the profile of the iron K$_{\alpha}$ (6.4 keV) line can 
be used to probe the structure of the accretion disk, because it is 
thought to result from fluorescence of the dense gas in the
geometrically thin and optically thick regions of the inner accretion 
disk ($\sim$ 10 $r_{\rm g}$). The most famous example is the spectrum of 
the Seyfert 1 (Sy-1) galaxy MCG--6-30-15, which shows a relativistically 
broadened Fe K$_{\alpha}$ emission line, first detected by $ASCA$ 
(\cite{tan95}).
Similar broad relativistic iron line profiles have
been detected in several other type-1 AGNs, although they are perhaps 
somewhat less common than anticipated from the $ASCA$ 
era (e.g., \cite{nan97}). In this context, studies of the  
iron line profile in $radio$-$loud$ AGN provides important clues to the 
disk-jet connection, particularly when compared with Sy-1s.

3C~120 ($z$ = 0.033) is the brightest broad line radio galaxy (BLRG), 
exhibiting characteristics intermediate between those of FR-I radio 
galaxies and BL Lacs. It has a one-sided superluminal jet on 100 kpc 
scales (\cite{wak87}),  and superluminal 
motion (with an apparent velocity $\beta_{\rm app}$ = 8.1) has been 
observed for the jet component. This provides an upper 
limit to the inclination angle of the jet to the line of sight of 
14 deg (\cite{era98}). Interestingly, \cite{mar02} 
found that dips in the X-ray emission of 
3C~120 are followed by ejections of bright superluminal knots 
in the radio jet, which clearly indicates an important connection 
between the jet and the accretion disk. In X-rays, 3C~120 has been 
known to be a bright ($\sim$ 5$\times$10$^{-11}$  
erg cm$^{-2}$ s$^{-1}$ at 2$-$10 keV), variable source with a canonical 
power-law spectral shape that softened as the source brightened (e.g., 
\cite{mar91}). A broad iron K$_{\alpha}$
line was first detected by $ASCA$ in 1994, with its width 
$\sigma$ = 0.8 keV and EW $\sim$400 eV (\cite{gra97}). 
Most recently, 3C~120 was observed for nearly a full orbit (130 ksec) 
with $XMM$-$Newton$ on 26$-$27 August 2003 (\cite{bal04, ogl05}). 
This clearly confirmed the presence of the neutral Fe line 
emission (57$\pm$7 eV in EW), which was slightly broadened with 
a FWHM of $\sigma$ = 9000$\pm$3000 km s$^{-1}$.  
Both of these papers  argued that the line 
profile is rather symmetric and no evidence was found for  
relativistic broadening, or alternatively the line arises from an accretion 
disk radius of $\ge$75 $r_g$ at an inclination angle of $\sim$10 deg
(where relativistic gravitational effects are almost negligible). 

3C~120 was observed with $Suzaku$ 
four times in February and March 2006 with a total (requested) 
duration of 160 ksec (\cite{kat07}). It was in a relatively high state during 
the 1st 40 ksec observation ($\#1$; the summed count rates of 
4 XISs detectors was 15.94$\pm$0.01 counts s$^{-1}$), 
then its count rate dropped by $\sim$20$\%$ in the 2nd observation, 
and finally reached a minimum in the 4th observation 
($\#4$; 12.02$\pm$0.01 counts s$^{-1}$). As given in Figure 5 ($left$), 
$Suzaku$ has successfully resolved the iron K line complex of 3C~120 
and also was first to verify the broad component's asymmetry. 
We showed that the iron line complex is composed of 
(1) a relatively narrow, neutral iron K line core (6.4 keV),   
(2) broad iron line emission possibly emitted from 
the accretion disk, and (3) an ionized $\sim$6.9 keV line. 
The residuals present after subtracting the iron line core 
are poorly modeled either by adding a simple broad Gaussian or 
by adding the Compton shoulder of the K$_{\alpha}$ line. 
A significant red-tail below 6.4 keV (in the rest frame of source) 
favors diskline emission from the inner accretion disk of $r_{\rm in}$ 
$\simeq$ 8.6$^{+1.0}_{-0.6}$ $r_g$ (\cite{kat07}). 

If indeed the broad line really originates from the inner 
accretion disk, it provides important clues to jet formation 
in the accretion disk.
For example, \cite{ree07} have discovered a similar broad iron 
line in the $Suzaku$/$XMM$-$Newton$
 spectra  of MCG--5-23-16, which is thought to 
originate from inner accretion disk ($r_{\rm in}$ $\simeq$ 20 $r_g$). 
These observations may imply that both the radio-loud 3C~120 and the 
radio-quiet MCG--5-23-16 have similar accretion disk structure, 
in contrast to suggestions that the optically-thick accretion disk 
is truncated in 3C~120 to a hot, optically thin flow at a distance 
of  $r_{\rm in}$ $\sim$ 100 $r_g$ 
(\cite{era00, zdz01, bal04}).  
Observations of the iron line profiles in various other broad line 
radio galaxies are important for the systematic comparison between  
Seyferts and BLRGs. We are planning to submit further deep observations
of other BLRGs in the next $Suzaku$ observation program to test this. 

Finally, another interesting discovery made by $Suzaku$ is that the 
variable component in  3C~120 is much $steeper$ ($\Gamma$ $\simeq$ 2.7) 
than the power-law emission component reported in literature 
(1.6 $\le$ $\Gamma$ $\le$ 1.8).  
One interesting idea to account for this steep, variable emission 
is the beamed radiation from the jet, though this component 
contributes only $\sim$ 20$\%$ at most, of the Sy-1 like X-ray emission 
in 3C~120 (i.e., emitted from the disk and corona). 
The low inclination angle implies that 3C~120 may 
have some ``blazar-like'' characteristics, such as rapid 
X-ray variability or a non-thermal spectrum extending to the 
gamma-ray energy band. Future deep $Suzaku$ observations,  
as well as continuing VLBI monitoring coincident with X-ray 
monitoring (as the campaigns reported by \cite{mar02}), 
sensitive measurements with $GLAST$, in quite different states of 
source activity will be crucial to understanding the 
nature of 3C~120. Figure 5 ($right$) clearly indicates that many BLRGs, 
including 3C~120, can be detected with $GLAST$ at MeV$-$GeV energy 
band near future.

\begin{figure}
  \includegraphics[height=.28\textheight]{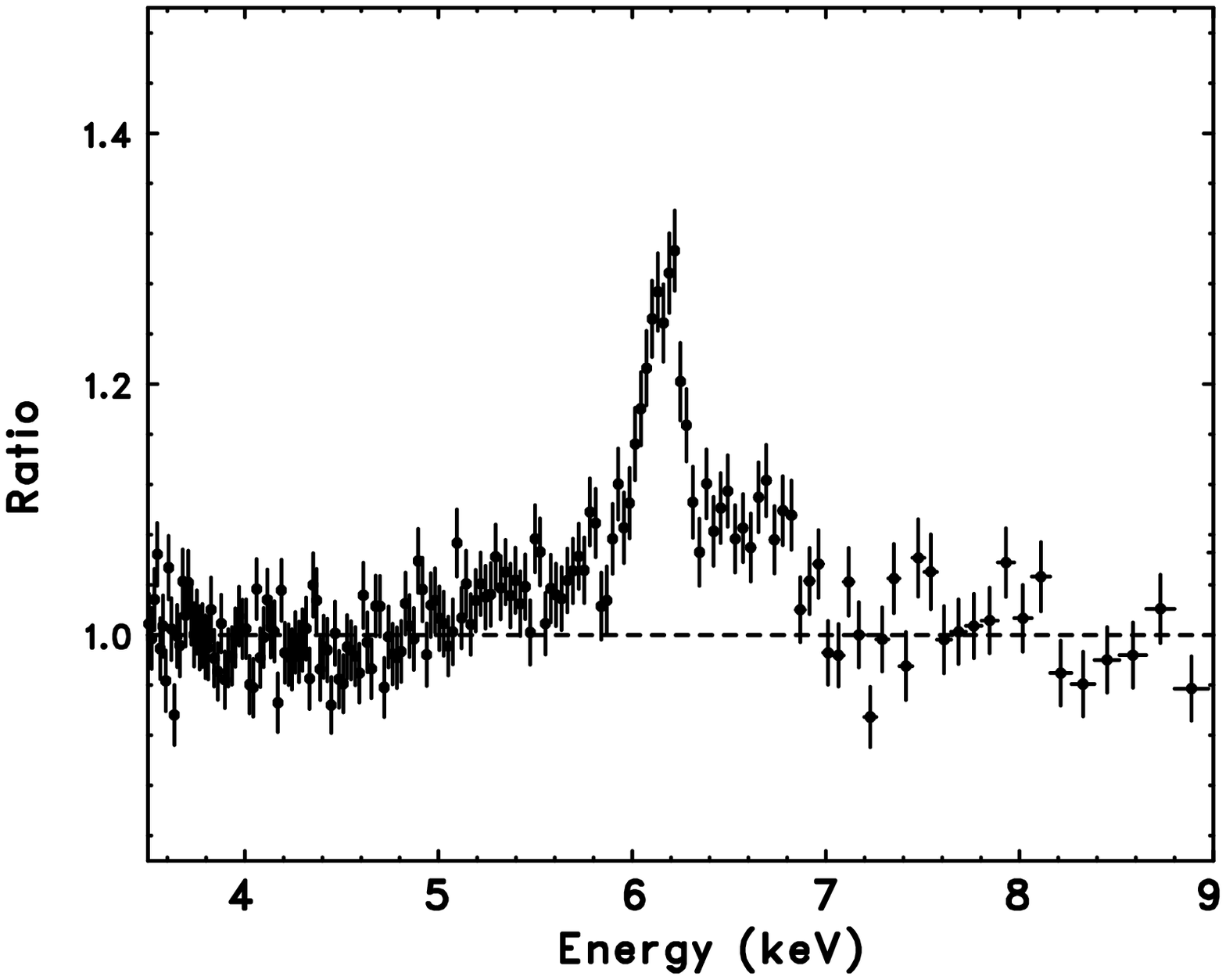}
  \includegraphics[height=.28\textheight]{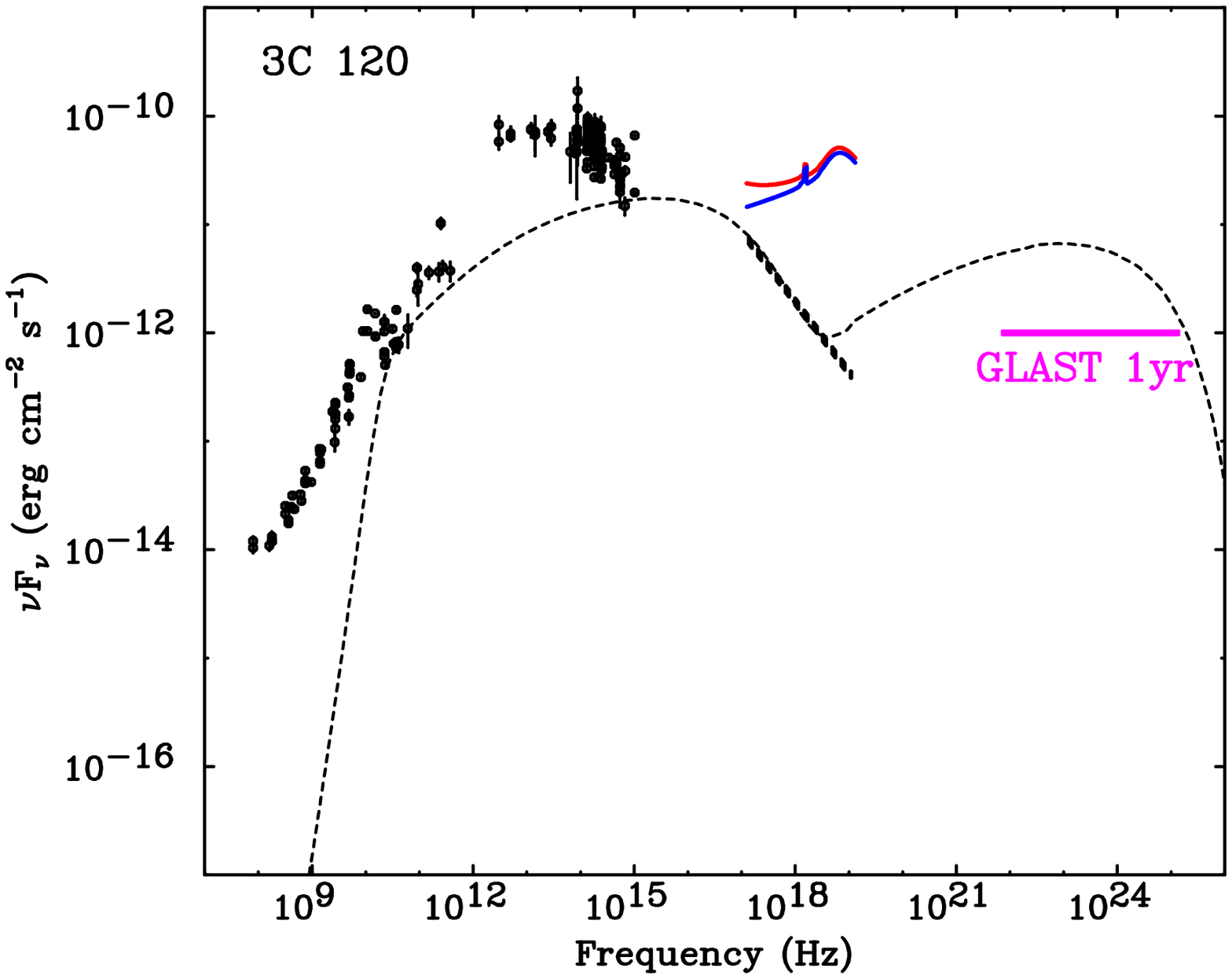}
  \caption{$left$ A close-up of the iron line profile of 3C~120, 
plotted as a ratio against a power-law of photon index $\Gamma$ = 1.7.
$right$: The spectral energy distribution of 3C~120. Red line shows 
the best-fit X-ray spectral model during the high stat, whereas blue 
lines shows that for the low state. Variable power-law component is 
shown as dashed line. The thin dotted line corresponds to an example fit
 of the one-zone homogeneous synchrotron self-Compton model as descrived
 in \cite{kat07}.}
\end{figure}

\section{4. Large Scale Jets: 3C~353}

The excellent spatial resolution of $Chandra$ X-ray Observatory has
opened a new era to study the large scale jets in powerful extragalactic radio
sources. Bright X-ray knots (so-called ``jet-knots'') are most often
detected, but the X-ray emissions from the hotspots and radio lobes 
are also reported in a number of FR II radio galaxies and quasars.
The X-ray emission observed from the extended lobes is well understood 
and modeled in terms of the
inverse Comptonization of the cosmic microwave background photons by the
low-energy electrons (IC/CMB; see the discussion in \cite{kat05,cro05}), 
providing strong evidence for
approximate energy equipartition between the radiating electrons and the
lobe magnetic field (with the particle pressure dominating over the
magnetic pressure by up to one order of magnitude). 

However, the most controversial issue is the origin of the intense
X-ray emission detected from knots in powerful quasar jets, such as
PKS~0637$-$752 or 3C~273 (\cite{sch00, mar01}, respectively). 
Here the X-ray knot spectra are much brighter than
expected from a simple extrapolation of the radio-to-optical
synchrotron continua, indicating that an additional or separate
spectral component dominates the jet's radiative output at high (X-ray)
photon energies. Very often, this emission is modeled in terms of
inverse-Comptonization of the CMB photon field by low-energy ($\gamma
\le 10^3$) electrons (\cite{tav00b,cel01}), 
which typically requires highly relativistic jet bulk
velocities of $\Gamma_{\rm jet} \ge 10$  even on kpc-Mpc scales. 
On the other hand, if there is significant beaming in 
powerful jets on large scales, then the detection of bright 
X-ray jet emission from FR~II radio galaxies, which are believed to 
be analogous systems to radio loud quasars but to be viewed 
with the jets at large angles to the line of sight, 
should be considered as unlikely. Such emission has, however, 
been detected in several objects (e.g., 3C~303, 3C~15, 
Pictor~A, or 3C~403; see \cite{kat03a,kat03b,har05,kra05}, 
respectively). Obviously, detection of
any X-ray $counterjet$ would be of primary importance in this respect,
since it would automatically exclude significant beaming, and thus
impose very severe constraints on the jet emission models. 

\begin{figure}
  \includegraphics[width=.5\textheight]{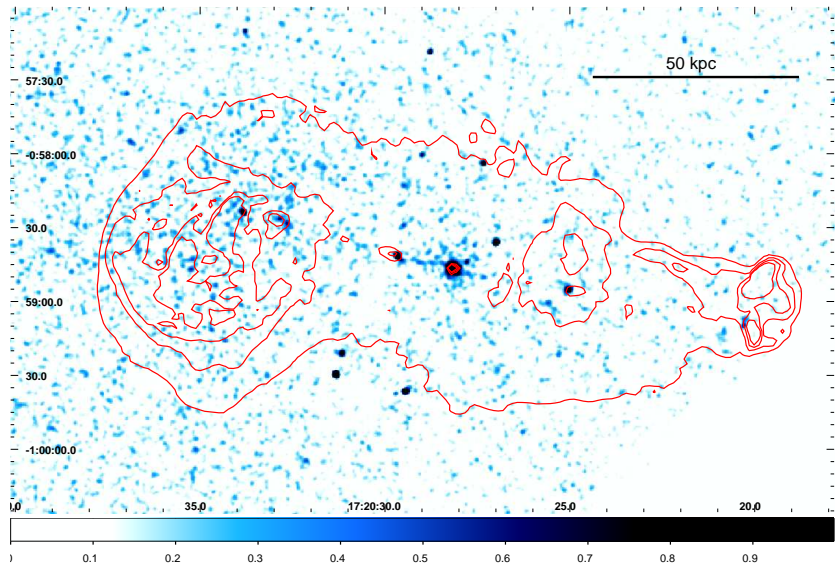}
  \caption{An X-ray image of 3C 353 ($Chandra$ 0.4$-$8.0 keV) 
overlaid with radio contours (VLA, 1.4 GHz). The contour levels are 
1.2, 4.6. 8.1 and 11.5 mJy beam$^{-1}$.}
\end{figure}
\begin{figure}
  \includegraphics[width=.5\textheight]{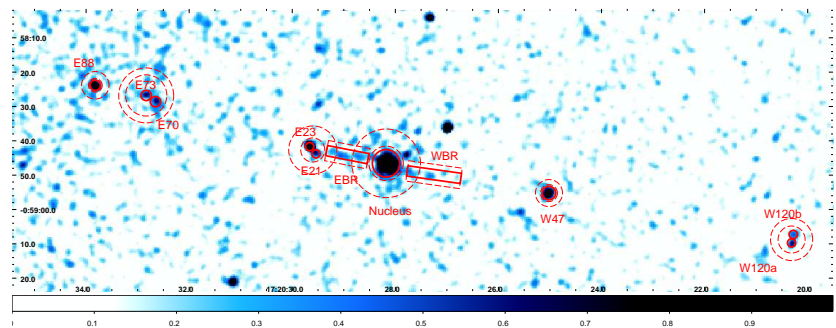}
  \caption{A $Chandra$ image of the central region of 3C~353.}
\end{figure}

3C~353 ($z = 0.0304$) is the fourth strongest radio source 
in the 3C catalog, with a total flux density $S_{\nu} \simeq
57$\,Jy at $1.4$\,GHz, and a projected size $\sim$ 4.5$'$. It
exhibits hotspots and a pair of large-scale FR~II-type jets, clearly
visible within filamentary lobes (\cite{swa98}). The jets in
3C~353, constituting about $1\%$ of the entire source luminosity, are
well collimated trains of knots with an average width $\simeq$
$4''$ and jet-counterjet radio brightness asymmetry $\simeq 2$
(\cite{swa96}). Both total and polarized intensity profiles across the
jets indicate that the bulk of the jet radio emission is produced at
the jet edges, and so presumably within a boundary shear layer. 
Since these jets are among only a very few FR~II jets wide enough 
to be resolved in X-rays, we planned and conducted a deep 
{\it Chandra} observation of 3C~353, in order to investigate the 
multiwavelength structure of powerful FR~II outflows. 
3C~353 was observed with {\it Chandra} in July 2007 with a total
(requested) duration of $90$\,ks (\cite{kat08b}). 
Figure 6 shows an exposure corrected image of 3C~353 in
the energy band $0.4-8.0$\,keV, with $1.4$\,GHz radio contours ($1.2$,
$4.6$, $8.1$, and $11.5$\,mJy\,beam$^{-1}$) overlaid. The X-ray image 
has been smoothed with a two-dimensional Gaussian function with
$\sigma=1.5$ pixels (1 {\it Chandra} pixel is $0.492''$). 
A zoom-up of the central region is separately given in  Figure 7.  
Most strikingly, the X-ray image clearly shows not only
the East (main) jet-knots, but also a bright knot in the West
counterjet (W47), that seems to be identified with the CJ2 radio knot
as given in \cite{swa98}. Also X-ray emission near the
West hotspot region  is (W120a,b) detected in the image. 

As discussed in \cite{swa98}, the width of the
radio jet in all these figures is much broader than the resolution of
the radio map ($1.3''$ for $1.4$\,GHz and $0.44''$ for $8.4$\,GHz).
The transverse profile in the radio cannot be well represented by a
simple Gaussian function, but instead shows a flat-topped profile
which is especially clear in W47. For quantitative comparison,
the width of the jet, at which the intensity becomes half of the
maximum (FWHM) are $3.43''/3.32''$ (W47), where the widths are 
quoted for the $1.4$\,GHz and $8.4$\,GHz radio maps respectively.
In contrast, the X-ray knots show no evidence for the 
flat-topped profiles seen in the radio images and are instead
well represented by a smooth Gaussian function, as shown in 
Figure 8 ($left$).  The observed FWHM 
of the X-ray jets are $1.79'' \pm 0.36''$ (W47), respectively, as
compared to the PSF width of $0.93'' \pm 0.01''$.  Therefore 
the X-ray \emph{knots} in 3C~353 are possibly narrower than their
radio counterparts (which are also situated further away from the
active nucleus), suggesting that the observed X-ray jet emission 
must be restricted to the central spine of the jet rather 
than to the jet boundary layer \cite{kat08b}.

\begin{figure}
  \includegraphics[height=.28\textheight]{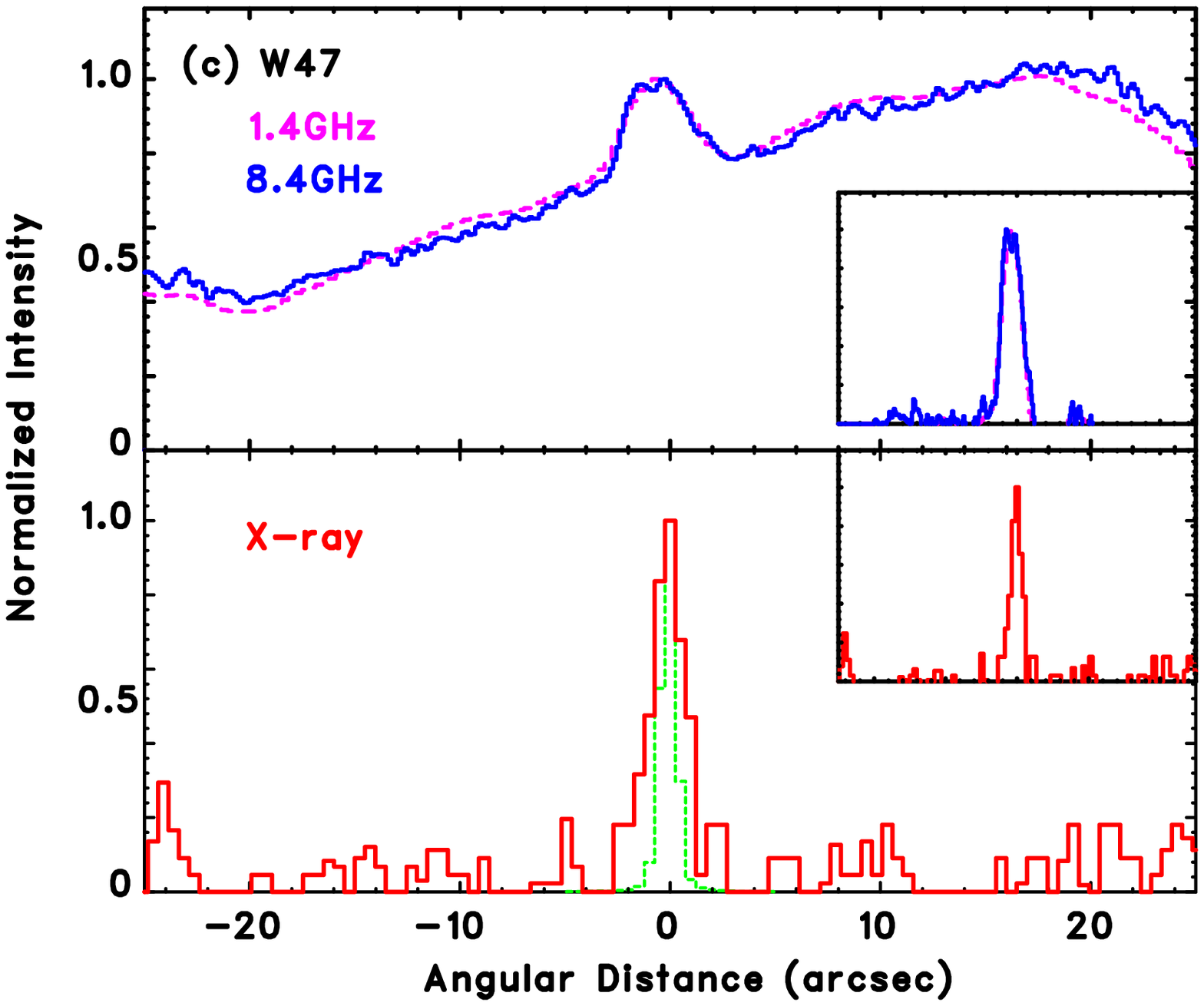}
  \includegraphics[height=.28\textheight]{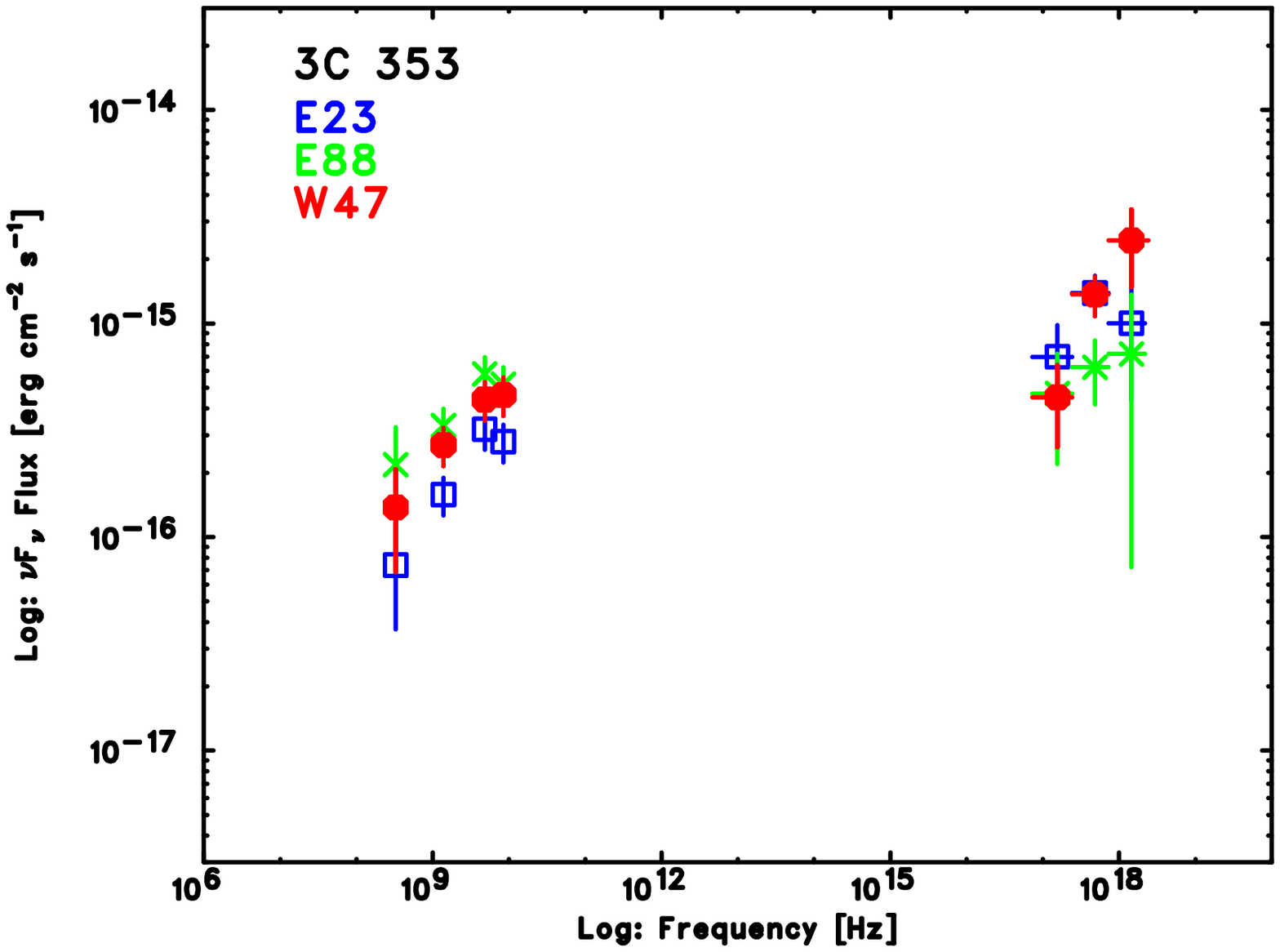}
  \caption{$left$: Comparison of transverse profile in the jet knot W47 
(counterjet). Above: the radio emission at 1.4 GHz (blue) and 8.4 
GHz (magenta). Below: the X-rays in the 0.4$-$8.0 leV range. The 
green dotted line shows the point spread function. $right$: Comparison 
of the SED of the jet-knots (E23, E88) and counterjet knot (W47).}
\end{figure}

Finally, we constructed the spectral energy distributions of the bright
jet knots E23, W47 and hotspot E88. Figure 8 ($right$) compares the spectral 
energy distributions, showing that the overall spectral 
features are remarkably similar to each other, suggesting that 
\emph{the same} physical process is at work for the X-ray 
production in the E23, E88 and W47 knots. One may note, however, 
that the X-ray spectral points cannot be connected smoothly 
with the extrapolation of the radio data, whether we assume either 
a single or a broken power-law form for the radio-to-X-ray continuum.
Note that \cite{swa98} constrained the jet inclination to the
line of sight in 3C~353 to be $60^{\circ} < \theta_{\rm j} <
90^{\circ}$; these large angles to the line of sight are strongly
supported by the observed two-sidedness of both the radio and X-ray
jets. Accordingly, we can approximately substitute $\delta_{\rm j}
\sim 1 / \Gamma_{\rm j}$, obtaining the observed luminosity 
ratio $ L_{\rm ic/cmb} / L_{\rm syn}$ independently of the
 jet kinematic factors. The resulting value, $\sim
10^{-3}$, is in strong disagreement with the observed X-ray-to-radio
luminosity ratio $L_{\rm X} / L_{\rm R} \ge 1$. 
Thus, we conclude that the IC/CMB model cannot explain the observed
X-ray emission of the 3C~353 jets, unless very large departures from
energy equipartition, $B \ll 100 \, \Gamma_{\rm j}^{-1}$\,$\mu$G, are
invoked.  

The only possibility left is therefore that the observed X-ray
emission of 3C~353 results from the synchrotron radiation of some
flat-spectrum high-energy electron population, most likely separate to
the one producing the observed radio emission. The required Lorentz
factors of the electrons emitting synchrotron photons with the
observed keV energies are $\gamma_{\rm X} \sim (\nu_{\rm keV} / 4.2
\times 10^6 \, B \, \delta_{\rm j})^{1/2} \sim 3 \times 10^7 \,
\Gamma_{\rm j}$ (assuming the scaling of the magnetic field as given
in equation 2 above and, again, $\delta_{\rm j} \sim 1 / \Gamma_{\rm
j}$). It has already been shown that stochastic acceleration processes
taking place in large-scale extragalactic outflows may easily account
for the production of electrons with these Lorentz factors
(e.g., \cite{sta02}) and they are already invoked to
explain observations of the hotspots of FR~II sources and the jets of
FR~Is.

\section{5. Future Challenges: $GLAST$}

It is widely expected that $GLAST$ will detect a large number 
(probably between 3,000 and 10,000) of extragalactic sources, 
most of which will be identified as blazars. Moreover, the LAT
large field-of-view combined with scanning mode will provide a 
very uniform exposure over the sky, allowing constant monitoring 
of all detected blazars and flare alerts to be issued. Apparently, 
contenmporaneous/simultaneous multiwavelength campaigns are 
essentially important for both ``EGRET blazars'' 
(i.e., well-established sources) as well as newly detected sources.
In X-ray, many observatories are already being actively prepared. 
For example, we are planning dedicated campaigns of 7 QHBs as a part  
of $Suzaku$-AO3 as listed in Table 1. Assuming a large flare 
as that observed for 3C~279 in 1991, $Suzaku$ can  determine the 
X-ray spectrum up to 300 keV with an unprecedented accuracy. 
Coordinated observations
between $GLAST$ and $Suzaku$ are crucial for further understanding
the nature of various types of blazars.

\begin{table}
\begin{tabular}{lrrrr}
\hline
\tablehead{1}{r}{b}{Source Name\\}
& \tablehead{1}{r}{b}{Redshift\\}
& \tablehead{1}{r}{b}{Class\\}
& \tablehead{1}{r}{b}{Flux (2-10 keV)\\10$^{-12}$ erg/cm$^2$/s}
& \tablehead{1}{r}{b}{Flux ($\ge$100 MeV)\\10$^{-5}$ ph/cm$^2$/s} \\
\hline
PKS 0208-512 & 1.00 & HPQ & 9.5    & 85.5$\pm$4.5\\
Q 0827+243   & 0.94 & LPQ & 4.8    & 24.9$\pm$3.9\\
PKS 1127-145 & 1.18 & LPQ & 11.0   & 38.3$\pm$8.0\\
PKS 1510-089 & 0.36 & LPQ & 10.0   & 18.0$\pm$3.8\\
3C~454.3 & 0.86 & HPQ & 11.0   & 53.7$\pm$4.0\\
3C~279 & 0.54 & HPQ & 13.0   & 89.0$\pm$3.2\\
PKS 0528+134 & 2.06 & LPQ & 30.0   & 60.0$\pm$3.0\\
PKS 2126-15 & 3.30 & LPQ & 12.0   & Non detection\\
\hline
\end{tabular}
\caption{A list of ``VIP'' blazars to be simultaneously observed with 
$GLAST$ and $Suzaku$ during in 2008/09 season.}
\label{tab:a}
\end{table}


In addition to classical gamma-ray blazars discussed above, we expect that
several nearby FR-I and FR-II radio galaxies can be
detectable by $GLAST$ as ``mis-aligned'' blazars. This is actually
expected if a unification scheme between blazars and radio galaxies
is indeed applicable. In fact, $Suzaku$ revealed that a power-law
continuum of Centaurus A extends up to 200 keV
without spectral break or reflection component (\cite{mar07}),
and the SED is well fit by synchrotron and inverse Compton model as usually
applied for blazars (\cite{chi01}).
Another interesting possibility is to detect broad line radio galaxies
(BLRGs) with $GLAST$, as we have discussed in $\S$3.  In fact, 
careful reanalysis of archival EGRET data reveals that one of the 
BLRGs (3C~111), which is analogous to 3C~120, may have been detected 
as a possible gamma-ray emitter (\cite{nan07}).  
Finally, we expect an important new area of investigation will be 
developed via ToO (Target of Opportunity) observations of 
AGNs by ``$GLAST$ trigger'' in next few years. 


\begin{theacknowledgments}
JK acknowledges all the $Suzaku$ members who helped us 
in analyzing the data, and all the $GLAST$-AGN members for 
careful planning of multiwavelength campaigns. 
JK thanks  support by JSPS KAKENHI (19204017/14GS0211).
\end{theacknowledgments}



\bibliographystyle{aipproc}   


\bibliography{sample}

\IfFileExists{\jobname.bbl}{}
 {\typeout{}
  \typeout{******************************************}
  \typeout{** Please run "bibtex \jobname" to optain}
  \typeout{** the bibliography and then re-run LaTeX}
  \typeout{** twice to fix the references!}
  \typeout{******************************************}
  \typeout{}
 }



\end{document}